\documentclass[11pt]{article}
\usepackage{latexsym}
\usepackage{amssymb}
\usepackage{epsf}
\usepackage{amsmath}
\usepackage{comment}
\usepackage[dvipdfmx]{graphicx}
\usepackage[dvipdfmx]{color}

\newcommand{\gsim}{\gtrsim}

\begin{document}
\pagestyle{empty}

\renewcommand{\thefootnote}{\fnsymbol{footnote}}

\vspace*{1cm}

\begin{center}

{\bf \LARGE Stimulated Emission of Dark Matter Axion from}

{\bf \LARGE Condensed Matter Excitations} 
\\

\vspace*{2cm}
{\large 
Naoto Yokoi$^{1}$\footnote{yokoi@imr.tohoku.ac.jp} and 
Eiji Saitoh$^{1, 2, 3}$\footnote{eizi@imr.tohoku.ac.jp}} \\
\vspace*{1.5cm}

{\it $^{1}$Institute for Materials Research, Tohoku University, Sendai 980-8577, Japan \\
$^{2}$WPI-Advanced Institute for Materials Research, Tohoku University, \\
Sendai 980-8577, Japan} \\
$^{3}$Advanced Science Research Center, Japan Atomic Energy Agency, Tokai 319-1195, Japan
\end{center}

\vspace*{3.5cm}

\begin{abstract}
{\normalsize We discuss a possible principle for detecting dark matter axions in galactic halos. 
If axions constitute a condensate in the Milky Way, 
stimulated emissions of the axions from a type of excitation in condensed matter can be detectable.  
We provide general mechanism for the dark matter emission, and, as a concrete example, 
an emission of dark matter axions from magnetic vortex strings in a type II superconductor 
are investigated along with possible experimental signatures.
}
\end{abstract} 

\newpage
\baselineskip=18pt
\setcounter{page}{2}
\pagestyle{plain}

\renewcommand{\thefootnote}{*\arabic{footnote}}
\setcounter{footnote}{0}

\section{Introduction}
Recent discovery of the Higgs particle revealed that the standard model of particle physics 
correctly describes the fundamental constituents of our universe, 
up to the energy scale $\sim 1\, \textrm{TeV}$ \cite{Agashe:2014kda}.  
However, the current standard cosmology strongly indicates the existence of 
additional constituents, which are known as the dark matter \cite{Kolb:1990vq, Weinberg:2008zzc}. 
In particular, observations of the rotation velocities of galaxies imply that 
there exists the dark matter halo around the Milky Way, whose mass density is given by 
\begin{eqnarray}
\rho_{DM}~ \simeq~ 3 \times 10^{14}~ [\textrm{eV}/c^2\, m^3]~ \simeq~ 
5 \times 10^{- 22}~ [\textrm{kg}/m^3] . 
\label{eq: mass density}
\end{eqnarray}
Although there are various candidates of the dark matter, 
including the weakly interacting massive particles (WIMPs), 
the origin and properties of the dark matter remain a mystery, so far.  
(See \cite{Jungman:1995df, Bertone:2004pz} for the current status of the dark matter research.) 

Among them, the axion is one of the promising candidates for the dark matter in our universe. 
The axion is a hypothetical elementary particle, which gives the most plausible solution 
to the strong CP puzzle in quantum chromodynamics (QCD) 
\cite{Peccei:1977hh, Peccei:1977ur}.  
Through various experiments and astrophysical observations for axion search, the properties of 
the axion, such as the mass and coupling strength, are severely constrained 
\cite{Agashe:2014kda, Kuster:2008zz, Graham:2015ouw}.  
For example, the typical mass consistent with the experimental constraints can be summarized as
\begin{eqnarray}
10^{-6}\, \left[\textrm{eV}/c^2\right]\, \lesssim\, m_{a}\, \lesssim\, 10^{-3}\, \left[\textrm{eV}/c^2\right], 
\label{eq: mass range}
\end{eqnarray}
for the invisible QCD axion models \cite{Kim:1979if, Shifman:1979if, Zhitnitsky:1980tq, Dine:1981rt}. 
This small mass distinguishes the axions from other WIMPs, 
whose masses are assumed to be $m_{\textrm{wimp}} \gsim 10^{9}\, [\textrm{eV}/c^{2}]$. 
In contrast to the standard WIMPs which are usually fermions, axions obey the Bose-Einstein statistics. 
In order to consistently explain the dark matter density in terms of the axion, 
a large number density of the bosonic axions is required in our galaxy, and 
the possibility of the Bose-Einstein condensation of dark matter axions 
due to such a large density has been discussed 
\cite{Sikivie:2009qn, Erken:2011dz, Saikawa:2012uk, Davidson:2014hfa, Guth:2014hsa}. 

In this paper, we discuss a possible consequence of the Bose-Einstein condensation of 
the dark matter axions: \textit{stimulated emissions of the axions}. 
Based on the coupling between the axion field and the electromagnetic field, 
we discuss the stimulated emissions of the axions from collective excitations in  
various condensed matter systems. 
In particular, as a concrete example, we investigate the stimulated emission from 
the magnetic vortex strings in type II superconductors, where a mobile vortex ensemble, such as vortex flow or vortex liquid, is realized near the critical temperature. 
The emission rate of the dark matter axions are estimated, 
and a possible experimental signature of the emission is discussed.

\section{Properties of Dark Matter Axions}
In this section, we briefly summarize the fundamental properties of the axion, in particular, 
the interaction between the axions and the electromagnetic fields. 
We also discuss the properties of the axion which make it to be a good candidate for the (cold) dark matter, 
with an emphasis on the difference from other dark matter candidates, such as WIMPs.  
 
\subsection{Coupling of Axion with Electromagnetic Field}
Since the axion is a (pseudo) Nambu-Goldstone boson \cite{Weinberg:1977ma, Wilczek:1977pj} 
of the spontaneous breaking of the Peccei-Quinn (PQ) symmetry \cite{Peccei:1977hh, Peccei:1977ur}, 
the axion field $\phi_{a}$ couples to other particles through its derivative, originated 
from a universal coupling,
\begin{eqnarray}
{\cal L}_{\textrm{NG}} \simeq j_{\textrm{PQ}}^{\mu} \left(\partial_{\mu}\phi_{a}\right) ,
\end{eqnarray}
where $j_{\textrm{PQ}}^{\mu}$ is the conserved current of the PQ-symmetry. 
Although the strength of this coupling is very small for the invisible axion models, 
various experiments to detect the axions using this type of coupling have been implemented so far \cite{Kuster:2008zz, Graham:2015ouw}.   
    
The axion has an additional coupling in the leading order with the electromagnetic (EM) field, 
which is a parity-odd coupling originated from the quantum anomaly of the PQ symmetry:
\begin{eqnarray}
{\cal L}_{\textrm{int}} = g_{0}\,\phi_{a}\,\vec{E}\cdot\vec{B} .
\label{eq: axion coupling}
\end{eqnarray}
In this paper, we focus on this axion coupling (\ref{eq: axion coupling}) and 
discuss its consequence.\footnote{The experiments for the axion detection using this coupling  
have also been performed \cite{Kuster:2008zz, Graham:2015ouw}.}  
With this coupling, the action for the axion field and EM-field is given by 
\begin{eqnarray}
S &=& \int\!\! d^4x~ \frac{1}{2} \left[\frac{\hbar^{2}}{c^2} \left(\frac{\partial \phi_{a}}{\partial t}\right)^2 
- \hbar^2 \left(\vec{\nabla} \phi_{a}\right)^2 - m_{a}^2 c^2 \phi_{a}^2\right] \nonumber \\
&+& \int\!\! d^4x~ \left[ \left( \frac{\varepsilon_{0}}{2}\vec{E}^2 - 
\frac{1}{2 \mu_{0}} \vec{B}^2 \right) + g_{0}\,\phi_{a}\,\vec{E}\cdot\vec{B} + \cdots \right] ,
\label{eq: original action}
\end{eqnarray}
where $m_{a}$ is the mass of the axion and the dots represent higher order terms 
with respect to the coupling $g_{0}$. 
The coupling strength $g_{0}$ is constrained to be very small from various experiments and astronomical observations \cite{Agashe:2014kda, Kuster:2008zz, Graham:2015ouw}, and the current bound is given by 
\begin{eqnarray}
 g_{0}~ \lesssim~ 10^{-10}~ [\textrm{GeV}^{-1}]~ \simeq~ 10^{-50}~ 
\left[\frac{\textrm{sec}~ C^2}{\sqrt{\textrm{kg}~ m}}\right],
\end{eqnarray}
where the first value is in the natural units, and the second one is in the SI units with $C$ being 
Coulomb for the electric charge. (See the Appendix for the conversion between these unit systems.) 
We optimistically take the value $g_{0} \simeq 10^{-10} [\textrm{GeV}^{-1}]$ in this paper.  
Due to this small value, the higher order terms with respect to $g_{0}$ can be completely neglected. 
Although the coupling (\ref{eq: axion coupling}) leads to the axion-photon scattering 
and the axion decay to two photons, the scattering and decay rates are very small and 
the lifetime of the axion is much longer than the current age of the universe. 
This indicates that the axions produced in the early universe with some mechanism, 
such as the misalignment of the vacuum angle and the cosmic string decay, 
can remain for a long time, and these relic axions can be promising candidates for the dark matter 
in our universe. 
Particularly, the production mechanisms are mainly non-thermal, and the axions, which are  
produced around the time of the QCD transition, behave as non-relativistic particles from the beginning. 
So, the axion is a good candidate for the cold dark matter \cite{Kolb:1990vq, Weinberg:2008zzc}.  

If the axion accounts for the major part of the dark matter of our universe, 
the abundance of the axion in our galaxy should explain the mass density (\ref{eq: mass density}),  
which is determined by the observation of the rotation curve of various galaxies 
\cite{Jungman:1995df, Bertone:2004pz}.
The velocity distribution of the dark matter in the galactic halo has also been discussed, and 
it is known that the violent relaxation by gravitational attraction \cite{LyndenBell:1966bi} leads to the Maxwellian distribution, where the root-mean-squared velocity is given by  
\begin{eqnarray}
v_{a} \simeq 2.7 \times 10^5\, [m/\textrm{sec}] \sim 10^{-3}\, c \, .
\label{eq: average velocity}
\end{eqnarray}
As noted above, since $v_{a}/c \sim 10^{-3} \ll 1$, the axions in our galaxy 
behave as non-relativistic particles. 
Although the mass of the axion depends on the models beyond the standard model of 
the elementary particles, the plausible mass range is discussed from the various viewpoints including
cosmology, astronomy and elementary particle physics \cite{Kuster:2008zz}. 
In particular, in order to explain the dark matter abundance in our universe, 
the expected mass should be in the range (\ref{eq: mass range}).  
If we take $m_{a} \simeq 10^{-6}\, [\textrm{eV}/c^2]$ as a typical value, 
we can calculate the number density $n_{a}$ of the dark matter axions in our galaxy,
\begin{eqnarray}
n_{a} = \frac{\rho_{DM}}{m_{a}} \simeq 3 \times 10^{20}\, \left[m^{-3}\right] .
\label{eq: number density}
\end{eqnarray}

\subsection{Non-Relativistic Limit of Axion Field} 
Since the axions behave as non-relativistic particles ($v_{a} \sim 10^{-3} c$), 
we can reasonably consider the non-relativistic limit 
of the axion field $\phi_{a}(x, t)$. At first, we separate the rest mass contribution from 
the time dependence of the axion field as
\begin{eqnarray}
\phi_{a}(x, t) = \frac{1}{\sqrt{2\, m_{a}}}\left( \varphi_{a}(x, t)\, e^{-i \frac{m_{a} c^2}{\hbar} t} + 
\varphi^{\dagger}_{a}(x, t)\, e^{i \frac{m_{a} c^2}{\hbar} t} \right) ,
\label{eq: NR axion field}
\end{eqnarray}
where $\phi_{a}^{\dagger} = \phi_{a}$ is satisfied for the real scalar. 
Inserting this ansatz into the axion part of the Lagrangian (\ref{eq: original action}), 
we obtain the following non-relativistic action  
\begin{eqnarray}
S^{NR}_{a} = \int\!\! d^4 x\, {\cal L}_{a}^{NR}  
= \int\!\! d^4 x\, \left[ i \hbar~ 
\varphi_{a}^{\dagger} \frac{\partial \varphi_{a}}{\partial t} - \frac{\hbar^2}{2\, m_{a}} |\vec{\nabla} \varphi_{a}|^2 \right] , 
\label{eq: NR axion action}
\end{eqnarray}
where the second derivative term with respect to time is discarded. 
This is nothing but the Lagrangian of the Schr$\ddot{\text{o}}$dinger field, which satisfies 
the non-relativistic Schr$\ddot{\text{o}}$dinger equation. The non-relativistic axion field 
$\varphi_{a}(x, t)$ becomes a complex scalar field, whose dimension is given by $[L^{-3/2}]$.  
Here, we define the rest mass frequency as
\begin{eqnarray}
\Omega~ \equiv~ \frac{m_{a} c^2}{\hbar}~ 
\simeq~ 1.5 \times 10^{9}\, [\textrm{sec}^{-1}] ,
\end{eqnarray}
for the typical mass $m_{a} \simeq 10^{-6}\, [\textrm{eV}/c^2]$.

In the non-relativistic limit, the axion coupling is also renormalized.
Inserting the non-relativistic ansatz of the axion field (\ref{eq: NR axion field}) into the original 
coupling (\ref{eq: axion coupling}), we obtain the axion coupling in the non-relativistic limit,  
\begin{eqnarray}
{\cal L}^{NR}_{\textrm{int}}  
= g\, \left(\varphi_{a} e^{-i \Omega t} + \varphi^{\dagger}_{a} e^{i \Omega t} \right) 
\vec{E}\cdot\vec{B} .
\label{eq: basic formula of axion coupling}
\end{eqnarray}
Here, the non-relativistic coupling strength is defined as 
\begin{eqnarray}
g~ \equiv~ \frac{g_{0}}{\sqrt{2\, m_{a}}}~ 
\simeq~ 10^{-29}\, \left[\frac{\textrm{sec}~ C^2}{\textrm{kg}~ \sqrt{m}}\right] , 
\end{eqnarray} 
where the typical value $m_{a} \sim 10^{-6} [\textrm{eV}/c^2]$ is used. 
This is the basic formula describing the interaction between the dark matter axions 
and excitations in condensed matter systems.

\subsection{Dark Matter Axion as Condensate}
Based on the discussion above, we are lead to a natural but 
somewhat surprising consequence: the dark matter axions can constitute a 
(Bose-Einstein (BE)) condensate. 
At first, the non-relativistic Lagrangian of the axion (\ref{eq: NR axion action}) essentially describes 
non-interacting particles obeying the Bose statistics.\footnote{The coupling strength with 
themselves and other particles are so small that it can be essentially ignored.}  
Given the large number density (\ref{eq: number density}) and applying 
the standard argument on the BE condensation to this axion system, 
the critical temperature is estimated by 
$k_{B} T_{c} \simeq \left(2 \pi \hbar^{2}/m_{a}\right) n_{a}^{2/3} \sim 10^{7} [\textrm{eV}]$. 
Furthermore, the thermal energy of dark matter axions, $k_{B} T_{a}$, is estimated from
the average velocity (\ref{eq: average velocity}) to be $k_{B} T_{a} 
\sim (1/2) m_{a} v_{a}^{2} \simeq 10^{-13} [\textrm{eV}]$. 
Therefore, if the dark matter axions are in thermal equilibrium in our galaxy, 
the axion field is naturally expected to be in a BE condensate phase at the 
extreme low temperature $T_{a}~ (\ll T_{c})$.  
The possibility of the thermalization of the dark matter axions by the small self-interaction 
and the gravitational interaction has been recently discussed in the literatures 
\cite{Sikivie:2009qn, Erken:2011dz, Saikawa:2012uk, Davidson:2014hfa, Guth:2014hsa}, 
and some consequences from the resulting BE condensation have been explored.  
In this paper,  we assume that the dark matter axions do thermalize and form the BE condensate 
in our galaxy.\footnote{It has been argued the possibility that miniclusters composed of axions 
have been formed through the galaxy formation process \cite{Hogan:1988mp, Kolb:1993hw}.  
(For the recent discussions, see \cite{Guth:2014hsa, Davidson:2016uok, Chavanis:2016dab}.) 
In this case, we assume that a significant fraction of axions have been thrown off due to the collisions and mergings of such miniclusters, and some fraction of axions forms a smooth halo distribution.}   

Assuming that the dark matter axions are in a condensed phase, 
we can estimate the occupation number ${\cal N}_{k}$ of the mode with the typical momentum  
$p_{a} \simeq m_{a} v_{a}  \sim 10^{-9}\, [\textrm{eV}/c]$.  
The corresponding thermal de Broglie wave 
length is given by $\lambda_{a} = 2 \pi \hbar/p_{a} \sim 10^3\, [m]$, which characterizes 
the quantum coherence scale of the BE condensation, 
as discussed in \cite{Guth:2014hsa}. Using this wave length, the typical occupation number 
at the mode with  $k_{a} = 2 \pi/\lambda_{a}$ 
becomes very large :
\begin{eqnarray}
{\cal N}_{k} \simeq n_{a}\, \lambda_{a}^{3} = \left(\frac{\rho_{DM}}{m_{a}}\right)  
\cdot \left(\frac{2 \pi \hbar}{m_{a} v_{a}}\right)^{3}\, \simeq\, 10^{29} .
\end{eqnarray}
As explained below, this large occupation number ${\cal N}_{k}$ gives an enhancement factor for 
the transition probability, which is analogous to the case of the stimulated emission of photons,  
used in the LASER. Note that the occupation number ${\cal N}_{k}$ is proportional to $m_{a}^{-4}$. 
 
In this condensed regime, the non-relativistic axion field $\varphi_{a}(x, t)$ behaves 
as a \textit{coherent classical field} in the leading approximation,
\begin{eqnarray}
\langle\, \left|\varphi_{a}\right|^{2}\, \rangle \simeq n_{a} ~ \Longrightarrow ~  
\varphi_{a} \simeq \varphi^{\dagger}_{a} \simeq \sqrt{n_{a}} \sim 10^{10}\, \left[m^{-3/2}\right]. 
\label{eq: field ansatz of condensation}
\end{eqnarray}
Using this condensed ansatz of the axion field, we can further rewrite the axion coupling in the following form:
\begin{eqnarray}
{\cal L}^{BE}_{\textrm{int}} = g \left(\varphi_{a}\,e^{-i \Omega t} + \varphi^{\dagger}_{a}\,e^{i \Omega t} \right) 
\vec{E}\cdot\vec{B} ~\simeq~ 2\,g\,\sqrt{n_{a}} \cos (\Omega t)\,\vec{E}\cdot\vec{B}. 
\label{eq: Condensed formula of axion coupling}
\end{eqnarray}
The factor $\sqrt{n_{a}}$ also leads to an enhancement of the coupling strength $g$, and 
gives a time-periodic coupling for collective excitations, as will be seen in the following.

\section{Excitations Coupled with Axions in Condensed Matter}
Based on the basic interaction (\ref{eq: Condensed formula of axion coupling}), 
three types of excitations in condensed matter systems can possibly couple with the dark matter axions.
\begin{enumerate}
\renewcommand{\labelenumi}{(\Alph{enumi})}
\item Excitations creating the electric field $\vec{E}$ under the external 
magnetic field $\vec{B}^{\textrm{ex}}$: \\
For concreteness, we consider the homogeneous external magnetic field 
in the $z$-direction, $\vec{B}^{\textrm{ex}} = (0, 0, B_{0})$. In the external magnetic field,  
the excitations creating the $z$-component of electric field $E_{z}$ can couple with the axions. 
In this case, the basic coupling (\ref{eq: Condensed formula of axion coupling}) approximately becomes
\begin{eqnarray}
{\cal L}^{BE}_{\textrm{int}} \simeq g\,B_{0}\left(\varphi_{a}\,e^{-i \Omega t} + \varphi^{\dagger}_{a}\,e^{i \Omega t} \right)\left(\gamma a + \gamma^{*} a^{\dagger}\right) .
\label{eq: axion coupling for type  A}
\end{eqnarray}
Here, the electric field associated with the excitation is given by 
$E_{z} =  \gamma a + \gamma^{*} a^{\dagger}$  in the linear approximation regime, 
where $a$ is the annihilation operator for a mode of 
the excitation and $\gamma$ is the corresponding coefficient. 
The possible candidates for such excitations in condensed matter systems are 
plasmons in (semi)condcutors, polaritons in insulators, and vortices in superconductors, and so on. 

\item Excitations creating the magnetic field $\vec{B}$ under the 
external electric field $\vec{E}^{\textrm{ex}}$:\\
Similarly, considering the external electric field in the $z$-direction, $\vec{E}^{\textrm{ex}} = (0, 0, E_{0})$, 
the excitation creating the $z$-component of magnetic field can couple with the axions.  
Then, the basic coupling (\ref{eq: Condensed formula of axion coupling}) is given by
\begin{eqnarray}
{\cal L}^{BE}_{\textrm{int}} \simeq g\,E_{0}\left(\varphi_{a}\,e^{-i \Omega t} + \varphi^{\dagger}_{a}\,e^{i \Omega t} \right)\left(\gamma a + \gamma^{*} a^{\dagger}\right) ,
\label{eq: axion coupling for type B}
\end{eqnarray}
where the magnetic field associated with the excitation is assumed to 
be $B_{z} =  \gamma a + \gamma^{*} a^{\dagger}$. The candidates are 
magnons and domain walls in magnetic systems, etc.   

\item Excitations which can create both the electric and magnetic field:\\ 
Recent studies on multiferroic and topological materials show that there exist 
the excitations creating both the electric field in response to an external magnetic field and
the magnetic field in response to an external electric field. 
Through the magnetoelectric effect, these excitations have the coupling with 
the axions in the form of (\ref{eq: axion coupling for type A}) or 
(\ref{eq: axion coupling for type B}), depending on the external field.   
The interesting candidates for such excitations are 
skyrmions in chiral-lattice magnets \cite{Nagaosa2013} and 
condensed matter analogues of the axions in a 
topological insulator with time-reversal breaking 
(so-called topological magnetic insulator) \cite{Essin:2008rq, Li:2009tca}.\footnote{A realization of 
axion electrodynamics is also discussed by using optical lattices \cite{Bermudez:2010da}.}  
\end{enumerate} 

In the following, we focus on the case of the excitations of type A, since the discussions are 
almost parallel for those of type B and C. 
In particular, as a concrete example and an interesting possibility, the magnetic vortex strings 
in type II superconductors will be discussed thoroughly.

\section{Axion-Vortex Coupling in Type II Superconductors}
In this section, we discuss the effects of the axion coupling (\ref{eq: basic formula of axion coupling}) 
in the dynamics of type II superconductors and, in particular, magnetic vortex strings. 

\subsection{Ginzburg-Landau Description of Superconductor and Magnetic Vortex}
We start with the Ginzburg-Landau (GL) equations, which give the effective description of
type II superconductors  \cite{tinkham1996introduction, kopnin2001theory}. 
\begin{eqnarray}
&& 0 ~=~ \frac{\hbar^2}{2 m^{*}}\left(\vec{\nabla} - i \frac{e^{*}}{\hbar}\vec{A}\right)^2 \psi - \alpha\,\psi - 
\beta\, |\psi|^2\,\psi , \label{eq: order parameter}\\
&& \frac{1}{\mu_{0}} \left(\vec{\nabla}\times \vec{B}\right) ~=~ i \frac{\hbar e^{*}}{2 m^{*}} 
\left[\left(\vec{\nabla} \psi^{\dagger}\right) \psi - \psi^{\dagger}\left(\vec{\nabla} \psi\right)\right]  - 
\frac{e^{*\,2}}{m^{*}} |\psi|^2 \vec{A} . \label{eq: vector potential}       
\end{eqnarray}
Here, we denote the order parameter of superconductors as $\psi$, and $e^{*}$ and $m^{*}$ are 
the electric charge and mass of the order parameter, respectively. $\alpha$ and $\beta$ correspond to
the parameters which can be calculated from the microscopic theory \`a la BCS.  

From the second equation (\ref{eq: vector potential}), we can read the supercurrent, 
\begin{eqnarray}
\vec{J}_{s} &=& i \frac{\hbar e^{*}}{2 m^{*}} \left\{\left(\vec{\nabla} \psi^{\dagger}\right) \psi - \psi^{\dagger}\left(\vec{\nabla} \psi\right)\right\}  - 
\frac{e^{*\,2}}{m^{*}} |\psi|^2 \vec{A} \nonumber \\
&=& \frac{\hbar e^{*}}{m^{*}} |f|^2 \left( \vec{\nabla} \theta - \frac{e^{*}}{\hbar} \vec{A} \right) ,
\label{eq: supercurrent}
\end{eqnarray}
where we insert the decomposition of the order parameter, 
$\psi(x) = f(x)\,e^{i \theta(x)}$, for the second equality.

As is well-known, the GL equations have the magnetic vortex solutions 
that are topologically stable \cite{Abrikosov:1956sx, Nielsen:1973cs}. 
For the later discussion, we briefly summarize 
the solitonic vortex solution. A straight single vortex, which is stretched in the $z$-direction, is described 
by the field configuration with the cylindrical symmetry:
\begin{eqnarray}
\vec{A}_{\textrm{v}}(x) = \left(\frac{\hbar}{e^{*}}\right) \left(\frac{\vec{d}_{z} \times \vec{r}}{r}\right) A(r), 
\quad \psi_{\textrm{v}}(x) = \sqrt{\frac{|\alpha|}{\beta}}\, e^{i \theta} f(r) .  
\label{eq: vortex solution}
\end{eqnarray}
Here, $(r, \theta, z)$ are the cylindrical coordinates, and $\vec{d}_{z}$ is the unit vector in 
the $z$-direction. $A(r)$ and $f(r)$ are the functions of 
the radial coordinate $r$ only, and the flux is normalized to be the minimal flux quantum, $h/e^{*}$. 
With these ansatze, the GL equations are reduced to the following coupled ordinary differential 
equations with respect to $r$:
\begin{eqnarray}
0 &=& \xi^2 \left[\frac{1}{r} \frac{d}{dr}\left(r \frac{d  f(r)}{dr}\right) - 
\left(\frac{1}{r}-A(r)^2\right) f(r) \right] - f(r) + f(r)^3 ,  \label{eq: GL1}\\
0 &=& \frac{d}{dr}\left[\frac{1}{r} \frac{d}{dr}\Bigl(r A(r) \Bigr)\right] - \frac{\mu_{0}\, e^{*\,2} |\alpha|}{m^{*} \beta} 
\left(A(r) - \frac{1}{r}\right) f(r)^2 . \label{eq: GL2}
\end{eqnarray}
Here, we define the coherence length for the order parameter:
\begin{eqnarray}
\xi = \sqrt{\frac{\hbar^2}{2 m^{*} |\alpha|}} .
\end{eqnarray}
With the boundary conditions $f(r) \rightarrow 1$ and $A(r) \rightarrow 1/r$ at $r \rightarrow \infty$,\footnote{For non-singular solutions, the boundary condition $f(r) \rightarrow 0$ at 
$r \rightarrow 0$ is also imposed.}   
we can easily obtain the numerical solutions for these equations, which describe the straight vortex with 
minimal flux. Away from the vortex core ($f(r) \simeq 1$), we can obtain another characteristic length scale, 
\begin{eqnarray}
\lambda_{L} = \sqrt{\frac{m^{*} \beta}{\mu_{0}\, e^{*\,2} |\alpha|}} ,
\end{eqnarray}
which defines the penetration depth of the electromagnetic field in superconductors.

\subsection{Vortex Dynamics with Axion Coupling}     
Now, we consider the interaction between the axion fields $\varphi_{a}$ and 
magnetic vortex strings based on the axion coupling (\ref{eq: basic formula of axion coupling}). 
Since the vortex string is a typical example of the excitations of type A,  
we consider the external magnetic field $\vec{B}^{\textrm{ex}}$ in the $z$-direction, 
$B^{\textrm{ex}}_{z} = B_{0} = \textrm{const.}~ (B^{\textrm{ex}}_{x} = B^{\textrm{ex}}_{y} = 0)$. 
Then, the axion coupling becomes
\begin{eqnarray}
{\cal L}^{NR}_{\textrm{int}} \simeq g\, B_{0} \left(\varphi_{a}\, e^{-i \Omega t} + \varphi^{\dagger}_{a}\, 
e^{i \Omega t}\right) E_{z} . 
\label{eq: external axion coupling}
\end{eqnarray}
Here, we derive the electric field created by the vortex motion in a simple set-up. 
For concreteness, we consider the limit of large penetration depth and 
small coherence length, compared to the size of a superconducting sample. 
In this situation, an external magnetic field becomes almost homogeneous in the sample and
the core of magnetic vortex becomes infinitely thin.\footnote{In the small $\xi$ limit, 
we take the order parameter as $\psi(x) = f e^{i \theta(x)}$ with a constant $f$.}  Outside the core of vortex, 
the GL equation (\ref{eq: vector potential}) and supercurrent (\ref{eq: supercurrent}) give
$\vec{A} =  \left(\hbar/e^{*}\right) \vec{\nabla} \theta$.
From the AC Josephson effect and gauge invariance, 
an electrostatic potential is also given by $\Phi = - \left(\hbar/e^{*}\right) \dot{\theta}$. 
Since the motion of the vortex with (small) velocity $\vec{v}_{L}$ implies a time-dependent phase 
$\theta = \theta(\vec{x}-\vec{v}_{L} t)$, the electric field originated from the vortex motion is 
obtained as  
\begin{eqnarray}
\vec{E}_{\textrm{vor}} = - \vec{\nabla} \Phi - \frac{\partial \vec{A}}{\partial t} = 
\left(\frac{\hbar}{e^{*}}\right) 
\left(\vec{\nabla} \frac{\partial}{\partial t} - \frac{\partial}{\partial t} \vec{\nabla}\right) 
\theta(\vec{x} - \vec{v}_{L} t) = - \left(\frac{h}{e^{*}}\right) \left( \vec{v}_{L} \times \vec{d}\, \right) 
\delta^{2}(\vec{x}) , 
\label{eq: vortex electric field}
\end{eqnarray} 
where the multivalued property of the phase $\theta(x)$ is used, and $\vec{d}$ is a unit vector 
along the vortex axis, 
and the two-dimensional delta function has the support on the core of vortex.   

As seen from the vortex electric field (\ref{eq: vortex electric field}), the motion in    
the plane perpendicular to the vortex axis is important, and we focus on the 
perpendicular motion in the following. 
Concretely, we regard the vortex string as a point particle on the perpendicular plane 
with a mass $m_{\textrm{vor}}$, by integrating out the spatial coordinate dependence in the Lagrangian.   
Using the formula (\ref{eq: vortex electric field}) and the homogeneous property (\ref{eq: field ansatz of condensation}) of $\varphi_{a}$, 
the axion coupling can be written in the minimal coupling form, 
\begin{eqnarray}
L_{\textrm{int}} = \int\!\!d^{3} x~ {\cal L}^{NR}_{\textrm{int}} \simeq  
\left(\frac{g\, h\, \ell_{\textrm{vor}}}{e^{*}}\right) \left(\varphi_{a}\, 
e^{-i \Omega t} + \varphi^{\dagger}_{a}\, e^{i \Omega t}\right) \left(\vec{B}^{\textrm{ex}} \times 
\vec{d}\, \right) \cdot \vec{v}_{L} = \vec{{\cal A}}_{\textrm{eff}}\cdot \vec{v}_{L} ,
\label{eq: vortex interaction lagrangian}
\end{eqnarray} 
where $\ell_{\textrm{vor}}$ is the length of the vortex string. 
Here, the effective gauge field $\vec{{\cal A}}_{\textrm{eff}}$ is introduced and 
its magnitude is given by 
\begin{eqnarray}
\left|{\cal A}_{\textrm{eff}}\right| \simeq \left(\frac{g\, h\,\left|\varphi_{a}\right| B_{0}\, 
\ell_{\textrm{vor}}}{e^{*}}\right) . 
\end{eqnarray}  
The above analysis can be simply extended to the case of general type II superconductors, 
using the time-dependent Ginzburg-Landau theory \cite{schmid1966time, kopnin2001theory}. 

This term and a standard kinetic term lead to the Lagrangian of a charged particle on
the perpendicular plane, coupled with an AC electric field, 
\begin{eqnarray}
L_{p} = \frac{m_{\textrm{vor}}}{2}\,\vec{v}_{L}^{\, 2} + \vec{{\cal A}}_{\textrm{eff}}(t)\cdot\vec{v}_{L}\, .
\end{eqnarray} 
Thus, our problem on the perpendicular motion of a rigid vortex with the axion coupling 
(\ref{eq: external axion coupling}) is reduced to the problem of a charged particle 
coupled with an AC electric field, whose frequency is 
$\Omega \sim 1\, [\textrm{GHz}]$ and magnitude is given by
\begin{eqnarray}
\left|\vec{E}\right| \simeq \left(\frac{g\, h\, \Omega\, \sqrt{n_{a}}\, B_{0}\, \ell_{\textrm{vor}}}{e^{*}}\right) ,
\end{eqnarray}  
where we have used the ansatz of the BE condensation of the axion field 
(\ref{eq: field ansatz of condensation}). 
If we put the parameters $B_{0} \sim 10\, [\textrm{T}]$, $\ell_{\textrm{vor}} \sim 10^{-4}\, [m]$ 
as an example, the drift force by the effective ``electric field'' can be estimated as
$F_{\textrm{drift}} \sim 10^{-27}\, [\textrm{N}]$, which is so small that it is hardly detectable.   

Note that, for this coupling to work, we require the existence of the vortices 
or segments of the vortex which are non-parallel to the external magnetic field $\vec{B}^{\textrm{ext}}$, 
as is evident from (\ref{eq: vortex interaction lagrangian}).  
For this purpose, we consider the following types of configurations of the vortex strings:
\begin{enumerate}
\renewcommand{\labelenumi}{(\alph{enumi})}
\item Tilted vortex strings against a tuned external magnetic field, due to 
         a boundary condition of a small superconducting sample.
\item Wavy vortex strings with the non-paralell segments or the vortex strings with  
         the kink-like segments, due to the small tension near the phase transition. 
\end{enumerate}       
In the following, we discuss mainly the case of the tilted vortex strings of type (a). 
For the wavy or kink-like vortex strings, the calculations can be done in the same way, 
except for replacing the length and number of the vortex strings with those of the segments.

\subsection{Stimulated Emission of Axions from Vortices}
In this section, we estimate the emission probability of axions from a vortex motion, using 
the first order perturbation theory, \textit{i.e.} the Fermi's golden rule.  
In order to realize the stimulated emission process by using the coupling 
(\ref{eq: vortex interaction lagrangian}), which is originated from 
the axion coupling (\ref{eq: axion coupling}), 
we require that there exist the excited states 
of the vortex dynamics having the energy level $\Delta E \sim \hbar \Omega$, 
which enables the emission of the non-relativistic axions with the energy,  
$E_{a} \simeq m_{a} c^2 + \frac{m_{a}}{2} v_{a}^{2}$. 
Furthermore, as in the case of the LASER, the non-equilibrium situation 
between the vortex system and the axions is also required, 
and the enough number of states of the vortices should be excited from the ground state.
This implies that the vortex system should have a higher temperature than both 
the axion temperature $T_{a}$ and $T_{\textrm{gap}} = \Delta E/k_{B}$. 
Therefore, we make the following assumptions on the dynamics of vortices in type II superconductors:
\begin{enumerate}
\renewcommand{\theenumi}{\Roman{enumi}}
\renewcommand{\labelenumi}{(\theenumi)}
\item The energy spectrum of the vortex dynamics has the excitations of the order of 
         $\hbar \Omega \sim 10^{-6}\,\textrm{eV}$, which corresponds to 
         the energy of the condensed axions. 
\item The vortices in the superconductor are in thermal equilibrium of the temperature 
         $T_{SC}$, higher than $T_{a}$ and $T_{\textrm{gap}}$, 
         without considering the effect of the dark matter axions.
\end{enumerate} 
To realize the assumption (I), we consider the magnetic vortices in mobile vortex systems 
where the excitaions with such a small energy level can be realized and observed 
as a resonance in experiments with microwave around GHz.  
Such mobile vortices can be realized in vortex creep, vortex flow, and vortex liquid in a type of 
superconductors \cite{Blatter:1994zz}.  
For the assumption (II), we take the typical temperature of a mobile vortex ensemble in superconductors 
as $T_{SC} \sim 4\, K$,\footnote{The experiments on superconductors are often performed 
at the liquid Helium temperature, $4.2\, K$.}  where the thermal energy 
of the vortex is typically $k_{B} T_{SC} \simeq 10^{-4}\,\textrm{eV}$. 
It should be noted that the energy spectrum of the vortex dynamics in mobile vortex systems is
expected to be broad around $\Delta E$, and the emission can take place within a certain range 
of the axion mass.   

In the second quantized language, the non-relativistic axion fields $\varphi_{a}$ has the Fourier expansion,
\begin{eqnarray}
&&\varphi_{a}(x, t) = \sum_{k} \frac{1}{\sqrt{V_{a}}}\, a_{k}\,  e^{i \vec{k}\cdot\vec{x} - i \omega t} , 
\label{eq: axion Fourier expansion} \\
&&\left[a_{k}, a^{\dagger}_{k'}\right] = \delta_{k,\,k'}\, ,\quad  \left[a_{k}, a_{k'}\right] = 
\left[a^{\dagger}_{k}, a^{\dagger}_{k'}\right] = 0 .
\end{eqnarray}
Here, $V_{a}$ is the volume of a box where the axion field is defined, and we finally  
take the limit $V_{a} \rightarrow \infty$ (or cosmological scale) as usual.  
Inserting the Fourier expansion (\ref{eq: axion Fourier expansion}), the axion coupling becomes
\begin{eqnarray}
&&\int\!\!d^{3}x\, g \left(\varphi_{a} e^{-i \Omega t} + \varphi_{a}^{\dagger} 
e^{i \Omega t}\right) \vec{E}\cdot\vec{B} \nonumber \\
&&\simeq \left(\frac{g\, h\,B_{0}\, \ell_{\textrm{vor}}}{e^{*}\, \sqrt{V_{a}}}\right) 
\sum_{k\,\simeq\, k_{a}}\left(a_{k}\, e^{- i \Omega t} + a^{\dagger}_{k}\,
e^{i \Omega t}\right) \vec{n}_{A}\cdot\vec{v}_{L} \nonumber \\
&& \equiv~ \vec{{\cal A}}_{\textrm{eff}}(t) \cdot \vec{v}_{L} , 
\end{eqnarray}
where $\vec{n}_{A} = (\vec{B}^{\textrm{ex}} \times \vec{d}~)/ \left|B^{\textrm{ex}}\right|$ 
is a unit vector in the direction of the effective gauge field.     
Here, we used the low-energy approximation, $e^{i\vec{k}\cdot\vec{x}} \sim 1$ 
and $\Omega \gg \omega$, and the axion distribution, which is highly concentrated around 
$k \simeq k_{a}$ based on the assumption of the axion condensation. 

Since this coupling is the same form as the standard minimal coupling to the EM-field, 
the interaction Hamiltonian, which contributes the perturbative calculation, 
simply becomes
\begin{eqnarray}
H_{\textrm{int}}(t) = - \frac{\vec{{\cal A}}_{\textrm{eff}}(t)\cdot \vec{p}_{L}}{m_{\textrm{vor}}} .
\end{eqnarray}
From this interaction Hamiltonian, we can estimate the emission probability 
within the first-order perturbation, by using the Fermi's golden formula, 
\begin{eqnarray}
\textrm{Prob.}\, [\textrm{sec}^{-1}] = \int\frac{d^{3} k\, V_{a}}{\left(2 \pi\right)^{3}} 
\left(\frac{2 \pi}{\hbar}\right) \left|\langle i |H_{\textrm{int}}(\Omega)| 
f \rangle\right|^2 \delta(E_{i}-E_{f}\pm\hbar \Omega) ,
\label{eq: Fermi formula}
\end{eqnarray}
where the momentum integration is performed over the final states of 
the emitted axion.  
Since we assume that the axions form the BE condensate, 
the problem is rather simplified. 
In the condensed phase, the state can be expressed by a coherent state, 
which is an eigenstate of the annihilation operator:
\begin{eqnarray}
a_{k} |{\cal N}_{k}\rangle \propto \sqrt{{\cal N}_{k}}\, |{\cal N}_{k}\rangle, \quad 
\langle {\cal N}_{k} | {\cal N}_{k} \rangle = 1 .
\end{eqnarray}
From this property, the matrix elements for the axion sector is evaluated as
\begin{eqnarray}
\langle i|\, a_{k}^{\dagger}\,|f \rangle = \langle {\cal N}_{k}|\, a_{k}^{\dagger}\,| {\cal N}_{k} \rangle \simeq \sqrt{{\cal N}_{k}} ,
\end{eqnarray}
where ${\cal N}_{k}$ is the \textit{occupation number} of the axions with a condensed momentum $k_{a}$.

Inserting the explicit form of $\vec{{\cal A}}_{\textrm{eff}}(t)$ into the formula (\ref{eq: Fermi formula}) 
and taking the emission part proportional to $\delta(E_{i}-E_{f}-\hbar \Omega)$, the probability becomes 
\begin{eqnarray}
\textrm{Prob.}\, [\textrm{sec}^{-1}] = \left(\frac{g\, B_{0}\, \ell_{\textrm{vor}}}{e^{*}\, 
m_{\textrm{vor}}}\right)^{2} \hbar\, {\cal N}_{k} \int\!\!d^{3}k\,\left|\langle p_{L} \rangle\right|^2\, 
\delta(E_{i}-E_{f}-\hbar \Omega) ,
\end{eqnarray} 
where the matrix element for the vortex sector is denoted as 
$\langle p_{L} \rangle \equiv \langle i, \textrm{vor} |\,p_{L}\,| f, \textrm{vor}\rangle$. 
Since we assume that the temperature of the vortex system $T_{SC} \sim 4\, K$ is much higher than the temperature of the condensed axions $T_{a} \sim 10^{-9}\, K$, the dominant process is 
the axion emission to the condensate. Note that the (apparent) dependence of the volume $V_{a}$ 
is cancelled in the resulting probability formula.    

Further assuming the matrix element $\langle p_{L} \rangle$ is independent of 
the momentum of the emitted axion in the low-energy region, we can perform the momentum integration, 
\begin{eqnarray}
\int\!\!d^{3}k\, \delta(E_{i}-E_{f}-\hbar \Omega) = 4 \sqrt{2}\,\pi\,\frac{m_{a}^{3/2} 
\sqrt{\hbar \Omega}}{\hbar^{3}} .
\label{eq: axion DOS} 
\end{eqnarray}
Gathering the above formulas, we have the expression for the emission probability of the axion 
from a single vortex,
\begin{eqnarray}
\textrm{Prob.}\, [\textrm{sec}^{-1}] = {\cal N}_{k} \left(4 \sqrt{2}\,\pi\, m_{a}^{3/2} 
\sqrt{\hbar \Omega}\right)  
\left(\frac{g\, B_{0}}{\hbar\, e^{*}\, \rho_{\textrm{vor}}}\right)^{2} \left|\langle p_{L} \rangle\right|^2 ,
\end{eqnarray}
where we defined the vortex mass density $\rho_{\textrm{vor}} = 
m_{\textrm{vor}}/\ell_{\textrm{vor}}\, [\textrm{kg}/m]$.
 
Finally, in order to evaluate the matrix element $\left|\langle p_{L} \rangle\right|^2$, 
we consider a simple model of the vortex dynamics, for concreteness. 
The model is described by the Hamiltonian with an effective harmonic 
potential \cite{Blatter:1994zz}:\footnote{Another model 
can be considered based on the transverse oscillations of the vortex string with a small tension.}  
\begin{eqnarray}
H_{\textrm{vor}} = \frac{p_{L}^2}{2\, m_{\textrm{vor}}} + \frac{m_{\textrm{vor}}\, \Omega^2}{2} x^2 
= \hbar \Omega\, a^{\dagger} a ,
\end{eqnarray}
where the creation and annihilation operators, $\left[a, a^{\dagger}\right] = 1$, are introduced. 
According to our assumption (I), the energy level should be 
$\hbar \Omega \simeq m_{a} c^2 \sim 10^{-6}\, [\textrm{eV}]$.   
Using the relation $p_{L} = \sqrt{\frac{m_{\textrm{vor}}\,\hbar \Omega}{2}} 
\left( a + a^{\dagger} \right)$, the matrix element can be estimated as
$\left|\langle p_{L} \rangle\right|^2 \sim m_{\textrm{vor}} \left(\hbar \Omega\right)$.  
With this estimation, the emission probability from a single vortex (or a single segment of the vortex) 
becomes
\begin{eqnarray}
\textrm{Prob.}\, [\textrm{sec}^{-1}] \simeq {\cal N}_{k} \left(4 \sqrt{2}\,\pi\, 
m_{\textrm{vor}}\,\left(m_{a}\,\hbar \Omega\right)^{3/2}\right) \left(\frac{g\, B_{0}}{\hbar\, e^{*}\, \rho_{\textrm{vor}}}\right)^{2} .
\end{eqnarray}
If an experimental sample of a superconductor contains the multiple vortices, 
the probability is proportional to the number of the vortices (for the tilted vortices) or 
the number of segments (for the wavy or kink-like vortices), which are denoted as $N_{\textrm{vor}}$. 
Thus, the emission probability of the dark matter axions from the vortex ensemble in type II 
superconductors becomes
\begin{eqnarray}
\textrm{Prob.}\, [\textrm{sec}^{-1}] \simeq {\cal N}_{k}\, N_{\textrm{vor}} \left(4 \sqrt{2}\,\pi\, 
m_{\textrm{vor}}\,\left(m_{a}\,\hbar \Omega\right)^{3/2}\right) \left(\frac{g\, B_{0}}{\hbar\, e^{*}\, \rho_{\textrm{vor}}}\right)^{2} .
\end{eqnarray} 
As the typical values of an experimental set-up, using $B_{0} \sim 10\, [\textrm{T}]$, $\rho_{\textrm{vor}} \sim 10^{6}\, [\textrm{eV}/c^2\, m]$, 
$\ell_{\textrm{vor}} \sim 10^{-4}\, [m]$, and $N_{\textrm{vor}} \sim 10^{5}$, 
we can numerically estimate the probability, 
\begin{eqnarray}
\textrm{Prob.} \simeq 10^{10}\, [\textrm{sec}^{-1}] .
\end{eqnarray}

Our result implies that, in an appropriate circumstances, the emission probability of 
the dark matter axions can be sizable, even if the axion coupling strength $g_{0}$ is quite small. 
For this type of process, the Bose statistics and small mass of the dark matter axions are
crucial, which distinguishes from other dark matter candidates, such as WIMPs.  
We comment here the difference from the conventional conversion process of 
axions into photons with a background magnetic field through the axion coupling. 
In this process, a mode mixing between the coherent axion field and the EM-field is essential, 
where the both fields are treated as classical waves, and the aforementioned stimulated 
emission does not play any role \cite{Sikivie:1983ip, Adler:2008gk}. 
In our axion emission process, the vortex is a heavy localized object compared to the axion, 
and thus the spontaneous emission of axions can be discussed in analogy with the case of spontaneous 
radiation from an atom.  

\section{Possible Signature of Dark Matter Axions}
An immediate consequence from the stimulated emission of dark matter axions is
the axion-driven non-equilibrium states of the vortex ensemble in type II superconductors 
\cite{tinkham1996introduction, kopnin2001theory}. 
Emitted axions from the vortices are almost invisible, because the interactions with other particles 
are negligibly small. However, the vortices consequently lose thermal energy 
due to the emission of axions, by the amount, $E_{\textrm{lost}} \simeq \textrm{Prob.} \times m_{a} c^2\, [\textrm{eV}/\textrm{sec}]$.
This process lowers the (local) temperature of the vortex system, and leads to the non-equilibrium 
state in the vortex ensemble. It should be emphasized that the temperature of the 
condensate of the dark matter axions is extremely low, $T_{a} \sim 10^{-9} K$, and 
the superconductor with vortices has the higher temperature $T_{\textrm{SC}} 
\sim {\cal O}(1) K$. Thus, the coupled system of the axions and vortices cannot reach 
the equilibrium as a whole in the laboratory. 
One possible experimental signature of the axion-driven non-equilibrium can be 
detected by the temperature measurement of a superconducting sample 
with mobile vortices in the vortex flow or vortex liquid.  
If the stimulated emission discussed in the last section is indeed realized, 
the adiabatic superconducting sample lowers 
its temperature \textit{spontaneously}, whereas its temperature normally should rise 
in the absence of such an emission. 

Another possibility is to detect the transport phenomena of the magnetic vortices in 
type II superconductors, 
such as the vortex Hall effect from the vortex motions, in a 
non-equilibrium situation \cite{tinkham1996introduction, kopnin2001theory, Blatter:1994zz}. 

\subsection{Stimulated Emission from Other Excitations in Condensed Matter}
So far, we have intensively discussed the stimulated emission from the magnetic vortices in superconductors, 
and estimated the emission rates, as a concrete example. 
As discussed in Section 3, there are other types of excitations which can couple with 
the dark matter axions via the coupling (\ref{eq: Condensed formula of axion coupling}).  
If such excitations have the energy level $E = \hbar\Omega \sim 10^{-6}\, [\textrm{eV}]$, 
the same arguments as the vortex case can be applied straightforwardly.   

At first, we consider an excitation in type A with the external 
magnetic field $\vec{B}^{\textrm{ex}} = (0,\, 0,\, B_{0})$. 
In the linear approximation, the electric field created by such an excitation
is represented by 
$E_{z} = \gamma\, a + \gamma^{*} a^{\dagger}$, 
with the annihilation and creation operators of the excitation mode.\footnote{For the vortex case, 
$\gamma = \gamma^{*} = 
\left(\frac{h}{e^{*}\,\rho_{\textrm{vor}}}\right)\sqrt{\frac{m_{\textrm{vor}} \hbar \Omega}{2}}$.}   
In this case, the stimulated emission rate is generally given by the following formula
\begin{eqnarray}
\textrm{Prob.}\, [\textrm{sec}^{-1}] = {\cal N}_{k} N_{\textrm{exc}} \left(4 \sqrt{2}\,\pi\, m_{a}^{3/2} 
\sqrt{\hbar \Omega}\right) \left(\frac{g\, B_{0}\, |\gamma|}{2 \pi \hbar}\right)^2 .
\label{eq: general formula for type A}
\end{eqnarray}
Here, the number of the excitations in the sample is given by $N_{\textrm{exc}}$, and 
the excitation level is assumed to be $\hbar \Omega \simeq m_{a} c^2$.
For the excitations of type B, which create a magnetic field  
under the background electric field $\vec{E}^{\textrm{ext}} = (0,\, 0,\, E_{0})$, 
similar calculations based on the coupling (\ref{eq: axion coupling for type B}) can be applied, 
and we can obtain the emission rate, 
\begin{eqnarray}
\textrm{Prob.}\, [\textrm{sec}^{-1}] = {\cal N}_{k} N_{\textrm{exc}} \left(4 \sqrt{2}\,\pi\, m_{a}^{3/2} 
\sqrt{\hbar \Omega}\right) \left(\frac{g\, E_{0}\, |\gamma|}{2 \pi \hbar}\right)^2 ,
\label{eq: general formula for type B}
\end{eqnarray}
where $N_{\textrm{exc}}$ is the total number of the excitations in the sample.

If the excitations of type C in a multiferroic (or topological) material 
also have the energy spectrum of the order $10^{-6}\,[\textrm{eV}]$, 
the stimulated emission of axions can be realized with either an external electric field or  
a magnetic field via the coupling (\ref{eq: axion coupling for type A}) or 
(\ref{eq: axion coupling for type B}).  
The emission rate is given by the formulas (\ref{eq: general formula for type A}) or 
(\ref{eq: general formula for type B}) in the same way.

\section{Summary and Discussion}
We have discussed the coupling between dark matter axions and excitations in
various condensed matter systems, and physical consequences from the coupling. 
In particular, we investigated the stimulated emission of dark matter axions, 
which are expected to take place BE condensates in our galaxy, from the various excitations 
in condensed matter systems through the axion coupling.  
As a concrete example, the emission from magnetic vortex strings 
of mobile vortex systems in superconductors was investigated, and  
a possible experimental signature, that is the spontaneous cooling and 
the resulting non-equilibrium state of the vortex ensemble, was discussed.   
It is expected that the stimulated emissions possibly 
give a new strategy to detect dark matter axions. 
For more concrete experimental set-up, further understanding of 
dynamics of the vortex strings in the mobile vortex systems, such as vortex flow and vortex liquid states, 
is required, and the investigation of such vortex dynamics is a work in progress. 
Although we focused on the vortex strings in s-wave superconductors in this paper, 
the axion coupling and emission in the case of vortex strings in (chiral) p-wave supercondutors 
should also be interesting.

Generalizations to other dark matter models with axion-like particles are straightforward. 
The emission rates are proportional to $g_{0}^{2}/m_{a}^2$, where $g_{0}$ and $m_{a}$ are 
the coupling strength and mass of the dark matter axion, and 
are enhanced for smaller mass axion if condensed matter 
excitations have the energy spectrum $\hbar \Omega \sim m_{a} c^2.$\footnote{Recently, new detection methods of ultralight axion-like particles corresponding to BEC dark matters are proposed. 
See \cite{Kahn:2016aff, Aoki:2016kwl, Dev:2016hxv} for example.}  

There is another interesting problem: The excitation of type C, such as an analogue of 
the axion in a topological magnetic insulator, has the same effective coupling as the axion coupling. 
Thus the mixing between the dark matter axions and condensed matter axions
can occur in principle. Physical consequences from such a mixing will be discussed 
in a future work.  

From the perspective of thermodynamics, the axion condensate plays the role of 
heat bath of an extremely low temperature for the condensed matter excitations discussed in this paper. 
Therefore, the mechanism investigated in this paper can lead to a novel energy conversion 
process from the condensed matter excitations.

\noindent
{\large {\bf Acknowledgements}}

\noindent
The authors thank J. Lustikova, S. Murakami, K. Sato, Y. Shiomi, and F. Takahashi for useful discussions. 
This work was supported in part by Grant-in Aid for Scientific Research on 
Innovative Areas "Nano Spin Conversion Science" (26103005). 
The work of E. S. was supported in part by ERATO, JST. 

\appendix
\section{Dimensional Analysis on Axion Coupling}
In this Appendix, we determine the unit (or dimension) of the coupling strength $g_{0}$ between 
the axion and EM-field. In the following, we denote the dimension of 
the Mass, Length, and Time as $[M]$, $[L]$, and $[T]$,  respectively. 
For the units of the EM-field, we take the electric charge as the fourth element of our unit, 
and denote the dimension of charge as $[Q]$. Here, we take the MKSC units 
(C represents [Coulomb]).

Since the dimension of the action $S$ is given by $[S] = \left[M L^{2} T^{-1}\right]$, we can determine 
the dimension of $\phi_{a}$,
\begin{eqnarray}
[\phi_{a}] = \left[M^{-1/2} L^{-3/2}\right] .
\end{eqnarray} 
As is well-known, the dimensions of the EM-field are determined from the equations of motion for
the charged particles,
\begin{eqnarray}
[\vec{E}] = \left[M L T^{-2} Q^{-1}\right], \qquad 
[\vec{B}] = \left[M T^{-1} Q^{-1}\right] .
\end{eqnarray}
From these relations, we obtain the dimension of the vector potential, 
$[\vec{A}] = \left[M L T^{-1} Q^{-1}\right]$.

From the above dimensions, we have the dimension of the following term (operator), 
\begin{eqnarray}
[\phi_{a}\,\vec{E}\cdot\vec{B}] = \left[M^{3/2} L^{-1/2} T^{-3} Q^{-2}\right] .
\end{eqnarray}
Using the $[S] = \left[M L^{2} T^{-1}\right]$, we can determine the dimension of 
the axion coupling strength, 
\begin{eqnarray}
[g_{0}] = \left[M^{-1/2} L^{-1/2} T Q^{2}\right] .
\end{eqnarray}
Thus, the coupling strength $g_{\textrm{mks}}$ in the MKSC unit becomes
\begin{eqnarray}
g_{0} = g_{\textrm{mks}}~ \left[\frac{\textrm{sec}~ C^2}{\sqrt{\textrm{kg}~ m}}\right] .
\end{eqnarray}

The experimental bound (or constraint) of the value of  the axion coupling strength is 
usually discussed in the natural units over the communities of high-energy physics, 
cosmology and astrophysics. 
To convert the MKSC units into the natural units, the following conversion formulas are useful:
\begin{eqnarray}
1\, \textrm{kg} &=& 5.61 \times 10^{35}~ [\textrm{eV}/c^2] , \nonumber \\
1\, m &=& 5.07 \times 10^{6}~ [\hbar c/\textrm{eV}] , \nonumber \\
1\, \textrm{sec} &=& 1.52 \times 10^{15}~ [\hbar/\textrm{eV}] , \nonumber \\
1\, \textrm{eV} &=& 1.60 \times 10^{-19}~ [\textrm{Joule}] ,
\end{eqnarray}
where $c = 3.00 \times 10^{8}~m/\textrm{sec}$ is the light velocity.
 
In the natural units, we set $c=\hbar=\varepsilon_{0}=\mu_{0}=1$. The (dimensionless) fine structure constant 
$\alpha = e^2/4 \pi \varepsilon_{0} \hbar c = 7.30 \times 10^{-3}~ (\sim 1/137)$ gives 
the elementary electric charge, 
\begin{eqnarray}
e = \left(4 \pi \alpha\,\varepsilon_{0} \hbar c \right)^{1/2} [C],
\end{eqnarray}  
and $e=0.303$ (dimensionless) in the natural units. 

Thus, the axion coupling strength in the natural units $g_{\textrm{nat}}$ is given by
\begin{eqnarray}
g_{\textrm{nat}} \simeq 3.22 \times 10^{30} \times g_{\textrm{mks}}~ [\textrm{eV}^{-1}] .
\end{eqnarray}
The upper bounds of this coupling strength originate from various astronomical observations of sun, 
neutron stars, and supernovae, which lead to  
\begin{eqnarray}
g_{\textrm{nat}} \lesssim 10^{-19}\, [\textrm{eV}^{-1}] \quad \Longrightarrow \quad 
g_{\textrm{mks}} \lesssim 3 \times 10^{-50}~ \left[\frac{\textrm{sec}~ C^2}{\sqrt{\textrm{kg}~ m}}\right].
\end{eqnarray}  
This means quite tiny coupling strength.

\end{document}